\documentclass[conference, 10pt]{IEEEtran}
\usepackage{amssymb,amsmath}
\usepackage{setspace}

\usepackage{graphicx}
\usepackage{epsfig,cite,latexsym,subfigure}
\usepackage{theorem}
\usepackage{times}
\usepackage{epsf,cite,subfigure}
\usepackage{amsmath,amssymb}

\usepackage{floatflt}
\usepackage{url}
\usepackage{times}
\usepackage[usenames]{color}
\usepackage{dsfont}
\usepackage{mathptmx}
\usepackage[final]{pdfpages}
\newcommand{\set}[1]{\left[#1\right]}
\newcommand     {\paren}[1]{\left(#1\right)}

\newcommand{\eqnlabel}[1]{\label{eqn:#1}}
\newcommand{\eqnref}[1]{(\ref{eqn:#1})}

\title{\vspace{-2mm}{\Large Mitigation of Adversarial Examples in RF Deep Classifiers\\ Utilizing AutoEncoder Pre-training }}\vspace{-2mm}
\author{
\IEEEauthorblockN{\normalsize Silvija Kokalj-Filipovic, Rob Miller, Nicholas Chang, C. L. Lau} \\%\vspace{-2mm}\\
\IEEEauthorblockA{\small Perspecta Labs Inc. \\
\small\em \{skfilipovic, rmiller, nchang\}@perspectalabs.com}}%, jsamson@ltsnet.net}} 
\begin{document}
\maketitle
\begin{abstract}
Adversarial examples in machine learning for images are widely publicized and explored.  Illustrations of misclassifications caused by slightly perturbed inputs are abundant and commonly known (e.g., a picture of panda imperceptibly perturbed to fool the classifier into incorrectly labeling it as a gibbon). Similar attacks on deep learning (DL) for radio frequency (RF) signals and their mitigation strategies are scarcely addressed in the published work.  Yet, RF adversarial examples (AdExs) with minimal waveform perturbations can cause drastic, targeted misclassification results, particularly against spectrum sensing/survey applications (e.g. BPSK is mistaken for 8-PSK).  Our research on deep learning AdExs and proposed defense mechanisms are RF-centric, and incorporate physical-world, over-the-air (OTA) effects. We herein present defense mechanisms based on pre-training the target classifier using an autoencoder. Our results validate this approach as a viable mitigation method to subvert adversarial attacks against deep learning-based communications and radar sensing systems. %It remains unknown if the RF AdExs maintain their effects in the physical world, i.e., when AdExs are delivered over-the-air (OTA), and 
\end{abstract}
%\vspace{-3mm}
%\section*{}
\section{Intro}%\vspace{-1mm}
\begin{figure}[t] %FIGURE 00
\vspace{-1mm}
\begin{center}
%\hspace{-5mm} \includegraphics [width=3.5in]{goodfellow.eps} %sampling.eps 
\hspace{-5mm} \includegraphics [width=3.5in]{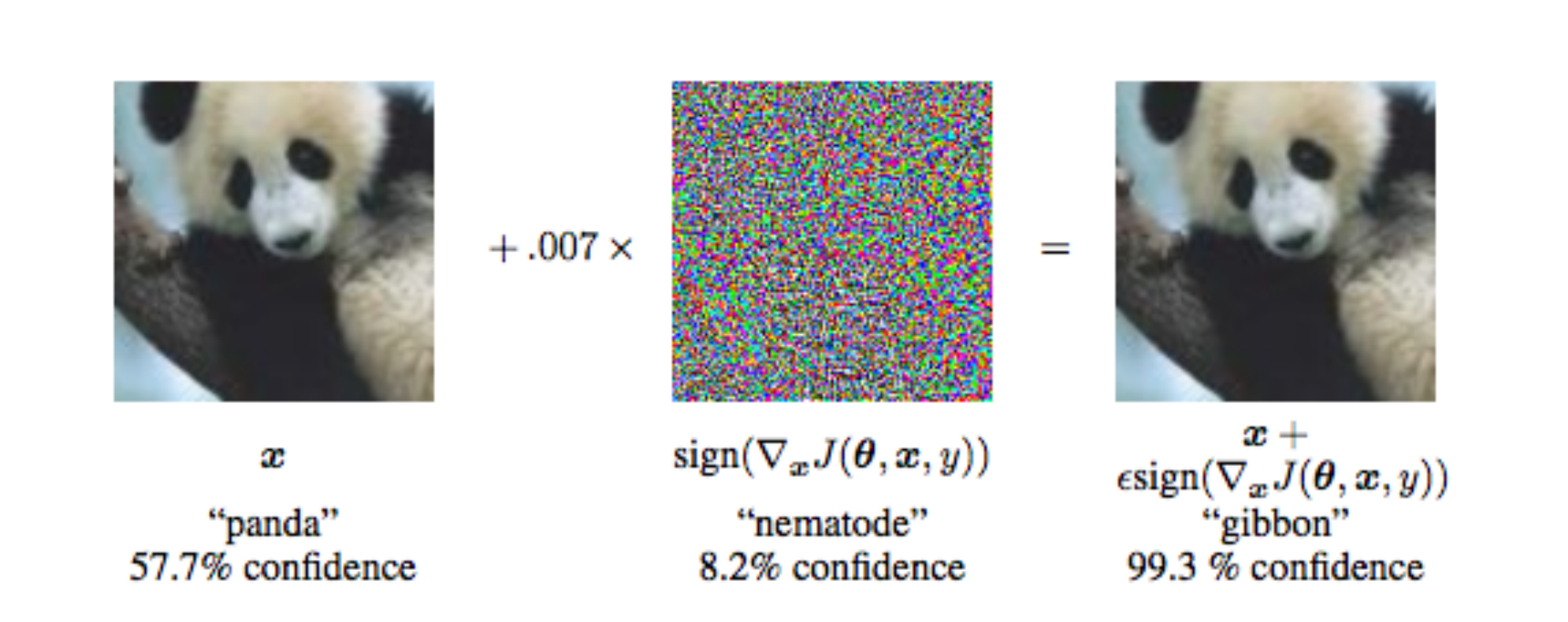} 
\caption{{Famous panda illustration of an adversarial image example against a DL classifier where a visually imperceptible, noise-like perturbation can fool the classifier to label it as gibbon}}\vspace{-4mm}   \label{fig:f00}
\end{center}
\end{figure}

A new research direction is emerging in the field of wireless communications, aiming to develop and evaluate deep learning (DL) approaches against classical detection and estimation methods in the radio frequency (RF) realm.  Spectrum sensing, especially in the context of cognitive radio, encompasses most of the radio signal detection problems that are being addressed.  The approach to DL in the RF domain differs greatly from the common current DL applications (e.g. image recognition, natural language processing) and requires special knowledge of RF signal processing and wireless communications and/or radar, depending on the signal utilization.  
%We have worked on several problems concerning the application of DL to cognitive radio, including modulation recognition, specific emitter recognition, protocol recognition, and network topology detection ().  We used synthesized or simulated wireless signals for modulation recognition and network topology recognition, and real signals collected over-the-air (OTA) for protocol recognition and specific emitter recognition. 
While research on adversarial examples in machine learning for images has been prolific, similar attacks on deep learning of radio frequency (RF) signals and the mitigation strategies are scarcely addressed in the published work, with only a couple of recent publications on RF \cite{Larsson, PAPRskfrm}. Adversarial examples (AdExs) are slightly perturbed inputs that are classified incorrectly by the Machine Learning (ML) model \cite{AdExSem1}. This perturbation is achieved by mathematical processing of the signal, e.g., by adding an incremental value in the direction of the classifier’s gradient with respect to the inputs (as in the FGSM attack illustrated in Fig.~\ref{fig:f0}~A), or by solving a constrained optimization problem. Popular deep learning (DL) models are even more vulnerable to AdExs as DL networks learn input-output mappings that are fairly discontinuous. Consider the images in Figure~\ref{fig:f00} \cite{AdExSem2}. The image on the left is the original image of a panda from the ImageNet dataset \cite{imagenet_cvpr09}, while the one on the right is derived from it by applying an FGSM attack of very low intensity. The  perturbation  of 0.007 added in the direction of the loss gradient corresponds here to the magnitude of the smallest bit of the normalized 8-bit RGB pixel encoding of the image. This is sufficient to elicit the GoogLeNet \cite{GoogleNet} to misclasify it as a gibbon. For further details about AdExs, please see seminal work, such as \cite{AdExSem1, AdExSem2}. Likewise for RF, adversarial examples can cause drastic, targeted misclassification results mostly in spectrum sensing/ survey applications (e.g. BPSK mistaken for 8-PSK) with minimal waveform perturbation.  However, it is not clear if the RF AdExs maintain their effects in the physical world, i.e., when AdExs are delivered over-the-air (OTA). Our research on deep learning AdExs and proposed defense mechanisms are RF-centric, and incorporate physical-world, OTA effects.  In this work we present defense mechanisms based on pre-training deep learning classifiers in the
RF domain by an autoencoder (AE) of the matching architecture.
%%%%%%%%%%%%%%%%%%%%%%%%%%%%%%%%%%%%%%%%%%%%%%%%%%%%%%%%%%%%%%
%%%%%%%%%%%%%%%%%%%%%%%%%%%%%%%%%%%%%%%%%%%%%%%%%%%%%%%%%%%%%%
\subsubsection{Existing Work}
The research in the area of RF-based DL of the PHY layer is still embryonic \cite{PHYDeep}. Modulation recognition \emph{(ModRec)} is the most popular application of DL here. Most of the existing work is based on convolutional (CNN) architectures \cite{ConvDeepRF}.  Paper \cite{OTADeepRF} features an in-depth study on the performance of DL ModRec methods on OTA captured RF communication signals synthetically designed in Software Defined Radio (SDR). The paper \cite{OTADeepRF} demonstrates that in the ModRec context DL provides significant performance benefits compared to conventional feature extraction methods. Apart from exploring optimal DL architectures and comparing their classification accuracy with state-of-the-art performance based on signal cumulants or their cyclo-stationary properties \cite{cyclostat}, this paper contributed a publicly available dataset  \cite{DeepSigDataset}, which we use here to demonstrate launching and mitigation of the attacks on DL that leverage adversarial examples of RF data points.

There are many methods to create AdExs. By definition, the following optimization problem describes the general approach:
\begin{align}
\nonumber &\min{||r_x||_p}\\
\nonumber s.t. &\ell(x) \neq \ell(x+r_x),\ \ and\\ 
& x+r_x \in \mathcal{X},
\end{align}
where $||\cdot||_p$ denotes the $l_p$ norm, and $\ell(x,\theta)$ is the decision rule by the NN with parameters $\theta$ evaluated at $x.$ The final constraint is somewhat arbitrary: it means that the adversarial example still belongs to the same space as the legitimate data point. Most current attacks are based on the gradient of a neural network’s loss function: White-box attacks use the target NN to compute the gradient; Black-box attacks use a surrogate network to approximate the gradient. 
We will be using Fast Gradient Sign Method (FGSM) \cite{AdExSem2} to illustrate our ideas. This attack takes the sign of the gradient and moves the data point $x$ one step in that direction.
\begin{align}
\nonumber &\hat{x} = x+\epsilon sign(\nabla_x J(x,y_{adv})),
\end{align}
where $x$ is the legitimate data point, and $\hat{x}$ is its adversarial example. $J(x,y_{adv})$ is the loss function for input $x$ evaluated for the targeted label. We denoted the targeted (adversarial classification) label as $y_{adv},$ which can be constant, or a random value, and the hyperparameter $\epsilon$ is usually a small number to limit the perturbation (note its value, $\epsilon = 0.07,$ in Figure~\ref{fig:f00}).

FGSM is a simple attack  and, at the same time, a basic principle used in iterative attack methods and constrained-optimization-based methods, hence representing a good reference for evaluation of new defense approaches. Some common iterative methods based on FGSM include: Basic iterative method (multiple steps of FGSM) \cite{AdExFGSMIter}, Carlini-Wagner method \cite{AdExCnW} (similar but modified objective function), Projected Gradient Descent (add noise, compute gradient, step, project back) \cite{AdExPGD}.
\section{Adversarial Examples: Problem Statement}
\begin{figure}[t] %FIGURE 1
\vspace{-1mm}
\begin{center}
%\hspace{-5mm} \includegraphics [width=3.4in]{conopsBud.eps} %sampling.eps 
\hspace{-5mm} \includegraphics [width=3.4in]{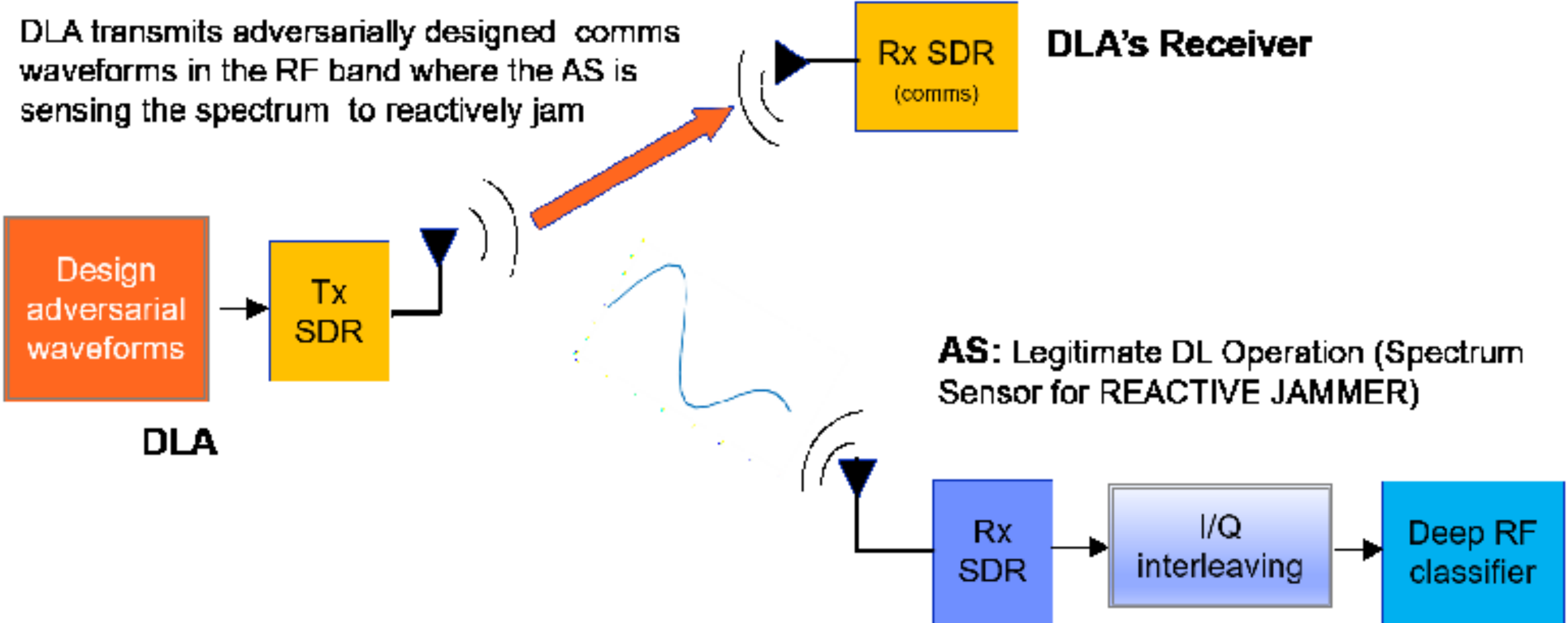}
\caption{{Motivating scenario for an OTA attack to DL classifier via RF AdExs}}\vspace{-4mm}   \label{fig:f1}
\end{center}
\end{figure}

Consider this scenario for an OTA attack on an RF DL classifier as a motivating example (Fig.~\ref{fig:f1}). The DL attacker (DLA) is at the transmitter of a communication system. Both DLA and the attacked system (AS) are using software defined radios (SDRs), although the DLA$'$s receiver does not have to be based on an SDR. The DLA$'$s goal is to elicit adversarial classification $y_{adv}$ at the AS on the signal $x$ designed based on a legitimate signal of class $y$. Despite the adversarial modifications targeted to elicit classification different from $y$, the designed AdEx needs to maintain high probability of being decoded as $y$ at the attacker$'$s own intended receiver, hence the perturbation of $x$ is constrained. 

The AS is sensing the spectrum in order to perform reactive jamming, e.g., to jam BPSK modulated waveforms that are part of traffic signaling (preambles, control-plane packets), which if corrupted makes the rest of communication meaningless.  Note that many preambles are BPSK-modulated, as well as the packets in the control plane of most protocols (such as acknowledgments), and if the they get corrupted by jamming the whole data packet is lost \cite{MillerVmimo}. Hence, the DLA would want to create the AdEx that disguises BPSK as QPSK (or other) to avoid the EW attack by the DL-based reactive jammer (i.e., AS) \cite{Wilhelm2011ReactiveJam}. Note that reactive jamming is very difficult to detect \cite{ReactXu05}, but it heavily relies on the inference based on spectrum sensing. If the inference is adversarially attacked, the jammer will be mitigated (by failing to reactively transmit a jamming signal). We here take the side of the jammer in attempting to detect such adversarial attacks.

In the case of images the {\em slight perturbation} applied to a legitimate data example is expressed as visual imperceptibility by a human viewer. RF adversarial examples is a nascent field, and as such the definition of imperceptible perturbation does not exist in the literature. 
For RF signals utilized for communications we define the imperceptible perturbation as any deformation of the RF waveform that can be filtered out by a receiver, e.g. via matched filters or correction codes, such that the bit-error rate is close to that of legitimate signals. An analogous definition can be made for radar using receiver operating characteristics (ROC). 

Mitigating adversarial inputs remains an open problem, even in the image-domain.  A complicating factor in detecting AdEx attacks include variations due to the physical world. Visual adversarial perturbations and their robustness given different backgrounds, lighting, and camera resolutions is discussed in \cite{Eykholt2017RobustPA}. The diversity of RF communications, radar, and spectrum sensing systems, and complex propagation channels makes this problem in the RF domain even more complex and unique. The effect of the channel (or interference) which is mitigated in well-designed communications receivers will persist at the DL classifier, thus changing the classification for both legitimate inputs and AdExs. Although we reflect on these issues, the proper consideration and modeling of the both hardware and the RF channel are outside the scope of this paper. %We focus on the uniqueness of RF data and its delivery to DL-based inference systems to defend against RF adversarial attacks. 

%%%%%%%%%%%%%%%%%%%%%%%%%%%%%%%%%%%%%%%%%%%%%%%%%%%%%%%%%%%%%%
%%%%%%%%%%%%%%%%%%%%%%%%%%%%%%%%%%%%%%%%%%%%%%%%%%%%%%%%%%%%%%

The defense mechanisms that we are proposing here rely on pre-training the DL classifier using an autoencoder. The idea of pre-training the classifier using an autoencoder is not new \cite{CompAEAdv} but our implementation and evaluation of this method is unique and crafted for RF signals. Some features of the RF waveform are corrupted during OTA delivery unintentionally, some are changed due to an adversarial attack. The autoencoder is expected to filter out non-salient features, which may lower the accuracy of classification, but also make it more robust to the adversarial and physical corruption. Common approaches to defense against adversarial attacks (mainly evaluated on image datasets) include: Gradient masking - hiding the gradient; Preprocessing - trying to “undo” the perturbations; Detecting AdExs - looking for distribution shifts; Certified methods - proving immunity to a set of perturbations. For details please see \cite{CertifiedDef} and references therein. Preprocessing is the closest of these strategies to the autoencoder-based method since the autoencoder operates by projecting the inputs into a space of lower dimensionality and filtering out the adversarial perturbations. 

Finally, adversarial training of neural networks is another common defense, and often complimentary to other methods.  It consists of generating the AdExs according to one or more attack methods, and retraining the NN with labeled  AdExs. We use adversarial training in conjunction with other mitigation and defense methods.

There are some complexities in DL of RF signals that we would like to highlight since our approach to solving those complexities impacts the presented results. Raw RF signal data is complex-valued, and traditionally split into the in-phase (I) and quadrature (Q) channels, resulting in a series of $I + jQ$ samples. Standard DL networks are not designed to handle complex-valued data, hence we must apply a transform to the real domain that preserves salient signal information. Our prior research leveraged expert feature transforms (e.g. FFTs, wavelets) to optimize performance by reducing the complexity (e.g. number of NN parameters) of the proceeding network. For this research we used interleaved I/Q samples of the DeepSig dataset comprised of synthetic data points from 24 modulation classes \cite{DeepSigDataset}. This simple transform from complex to real set, could be expressed as follows: for a data point $c$ of $k$ I/Q samples $c_1, \cdots, c_k$, where $c_{\ell} = I_{\ell} +j\ast Q_{\ell},$ the transformed vector has $2k$ real elements $\set{I_1, Q_1, I_2, Q_2, \cdots, I_k, Q_k}.$ For the Deepsig dataset that we used $k=1024,$ hence the input to the NN is a tensor of dimensions $\paren{\cdot, 2048}.$ Note that despite the conversion from complex data to the interleaved I/Q transform the accuracy of classifying four modulations represented by a subset of the DeepSig dataset gets close to 100\% if data points with $SNR \geq 14$~dB are used (see Figures~\ref{fig:f2} and \ref{fig:f2b}). Adversarial examples with $\epsilon = 0.1$ lower the accuracy by 30\% or more on average.  

%%%%%%%%%%%%%%%%%%%%%%%%%%%%%%%%%%%%%%%%%%%%%%%%%%%%%%%%%%%%%%
\begin{figure}[t] %FIGURE 0
\vspace{-1mm}
\begin{center}
\begin{tabular}{c c}
%\hspace{-10mm} \includegraphics [width=1.5in]{FGSM.eps} & \hspace{-2mm} \includegraphics [width=1.8in]{50-4-attacked01A4.eps}\vspace{-2mm}\\ 
\hspace{-10mm} \includegraphics [width=1.5in]{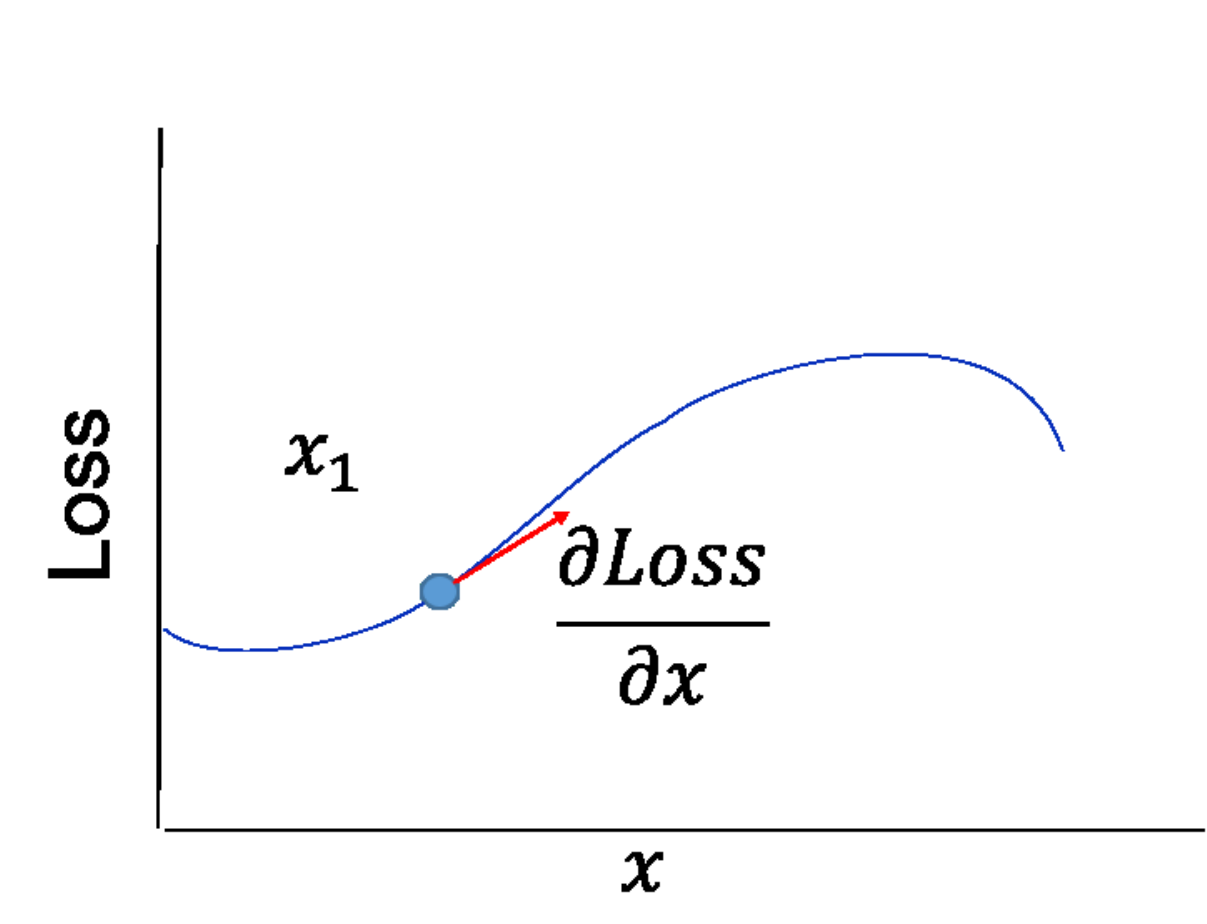} & \hspace{-2mm} \includegraphics [width=1.8in]{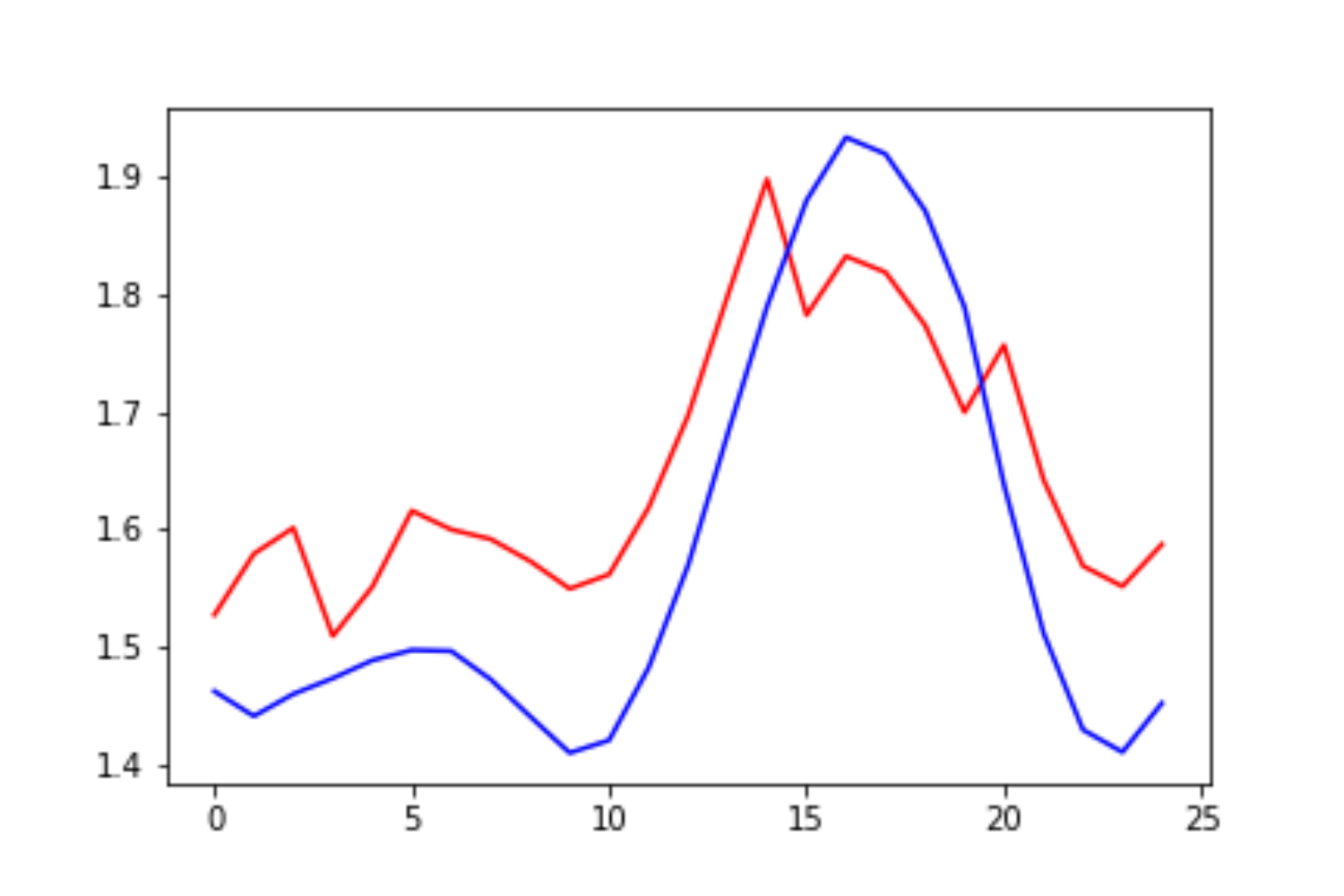}\vspace{-2mm}\\ 
{\small \textbf{A:} Basics of AdEx design}   & {\small \textbf{B:} Effect on a QPSK signal amplitude} %\vspace{-2mm}
\end{tabular}\\
%\hspace{-10mm}\includegraphics [width=3.1in]{BPSKadvI.eps}\vspace{-2mm}\\
\hspace{-10mm}\includegraphics [width=3.1in]{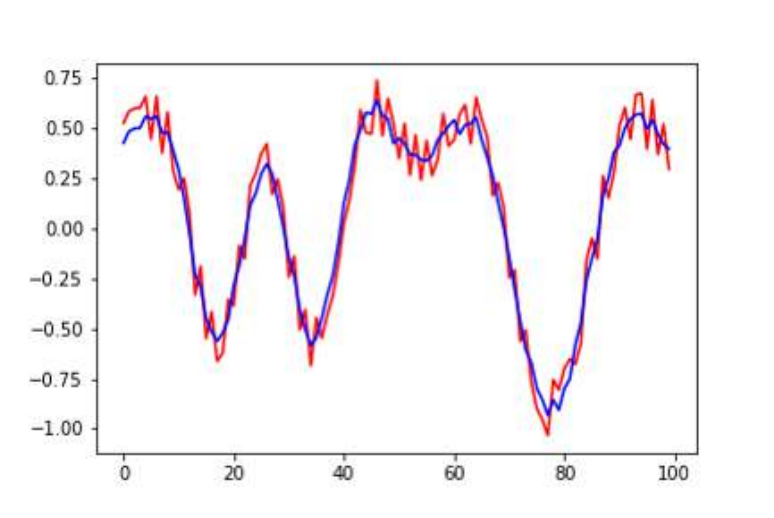}\vspace{-2mm}\\
\hspace{-6mm}{\small \textbf{C:} Effect on a QPSK data point (in-phase)}\vspace{-2mm}
\caption{{FGSM Attack (A) and its effect on a modulated RF signal (B) and its data point (C) }}\vspace{-2mm}   \label{fig:f0}
\end{center}
\end{figure}
The bottom (C) of Figure~\ref{fig:f0} shows 100 in-phase samples of a QPSK adversarial example (red) and its legitimate counterpart (blue). Similar effect is observed for the quadrature component. Neither I nor Q samples visually change much for a small perturbation $\epsilon$ (FGSM with $\epsilon = 0.1$). However, the modification induced on the signal amplitude (top right - B) is more pronounced, and depending on the $\epsilon$ value this may have other effects at the receiver. Note that the B plot shows 25 amplitude samples (from 50 samples of the data point).
\section{Proposed Method for Mitigation of Adversarial Examples}
\begin{figure}[t] %FIGURE 2
%\vspace{-1mm}
\begin{center}
%\hspace{-2mm} \includegraphics [width=3.5in]{convarch.eps} %sampling.eps 
\hspace{-2mm} \includegraphics [width=3.5in]{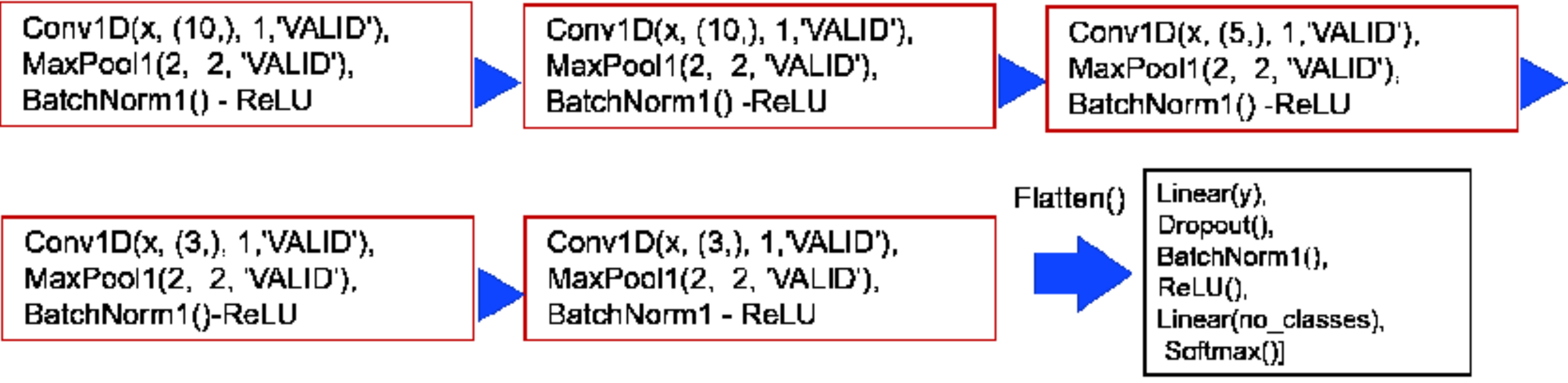}
\caption{{Architecture of the 1D-CNN classifier that was both classically trained and pre-trained by autoencoder; x = 4, y = 256}}\vspace{-4mm}   \label{fig:f2a}
\end{center}
\end{figure}

All the results presented here are based on a subset of the DeepSig dataset  – specifically BPSK, QPSK, 8PSK and 16QAM (DeepSig classes 3, 4 ,5, 12). We applied the FGSM attack using the CleverHans library (12). We compared results with and without the attack using Auto-encoder (AE) based training and conventional training of the convolutional neural network (CNN) presented in Figure~\ref{fig:f2a}.
%%%%%%%%%%%%%%%%%%%%%%%%%%%%%%%%%%%%%%%%%%%%%%%%%%%%%%%%%%%%%%
%%%%%%%%%%%%%%%%%%%%%%%%%%%%%%%%%%%%%%%%%%%%%%%%%%%%%%%%%%%%%%%

%%%%%%%%%%%%%%%%%%%%%%%%%%%%%%%%%%%%%%%%%%%%%%%%%%%%%%%%%%%%%%
\begin{figure}[t] %FIGURE 2
\vspace{-1mm}
\begin{center}
%\hspace{-5mm} \includegraphics [width=3.1in]{acc11b.eps} %sampling.eps 
\hspace{-5mm} \includegraphics [width=3.1in]{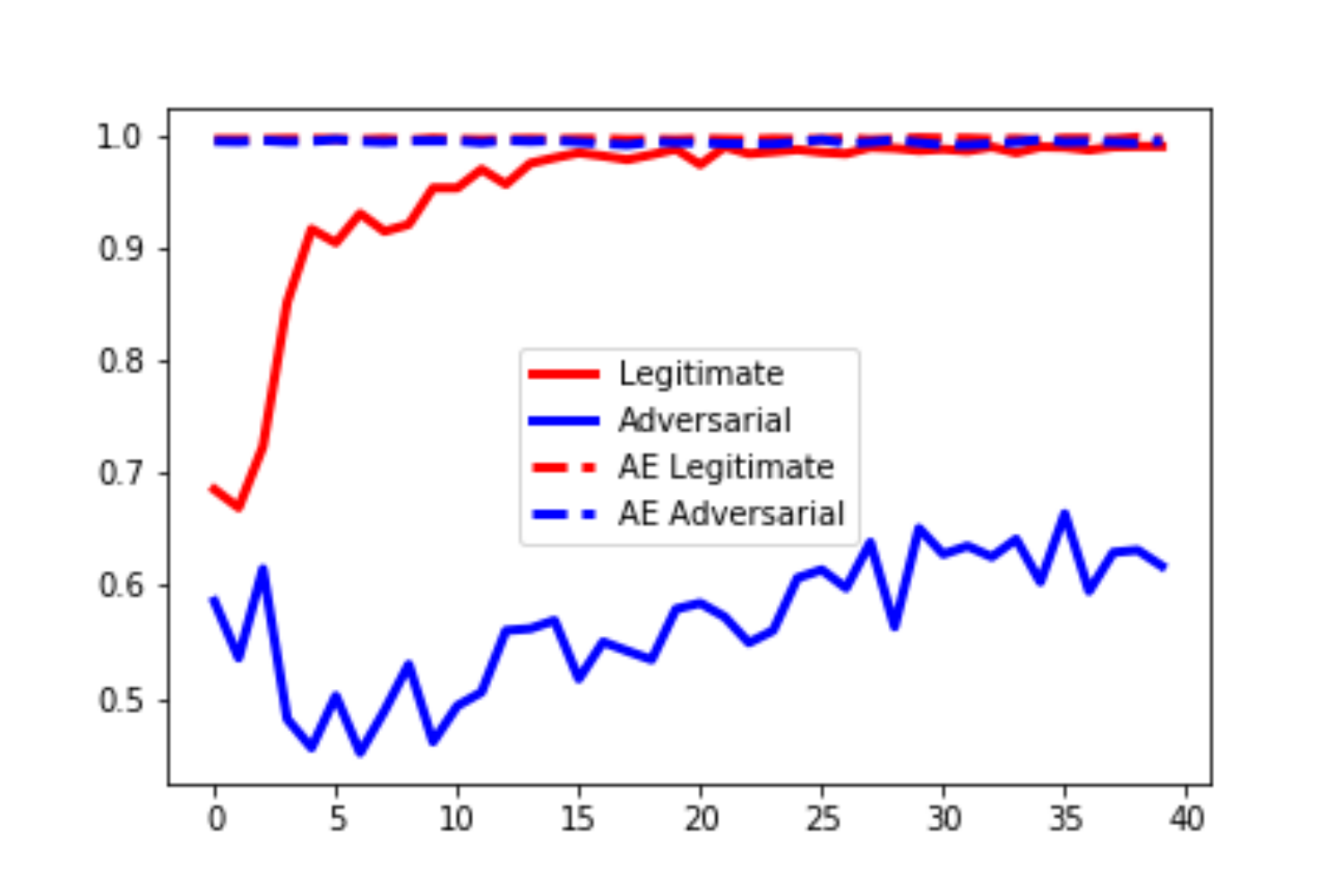} 
\caption{{Accuracy (over training epochs) of a 1D-CNN classifier when classically trained, vs pre-trained by autoencoder, for both legitimate data and FGSM AdExs}}\vspace{-4mm}   \label{fig:f2}
\end{center}
\end{figure}
%%%%%%%%%%%%%%%%%%%%%%%%%%%%%%%%%%%%%%%%%%%%%%%%%%%%%%%%%%%%%%%

%%%%%%%%%%%%%%%%%%%%%%%%%%%%%%%%%%%%%%%%%%%%%%%%%%%%%%%%%%%%%%
\begin{figure}[t] %FIGURE 2
\vspace{-1mm}
\begin{center}
%\hspace{-5mm} \includegraphics [width=3.4in]{AEtrain.eps} %sampling.eps 
\hspace{-5mm} \includegraphics [width=3.4in]{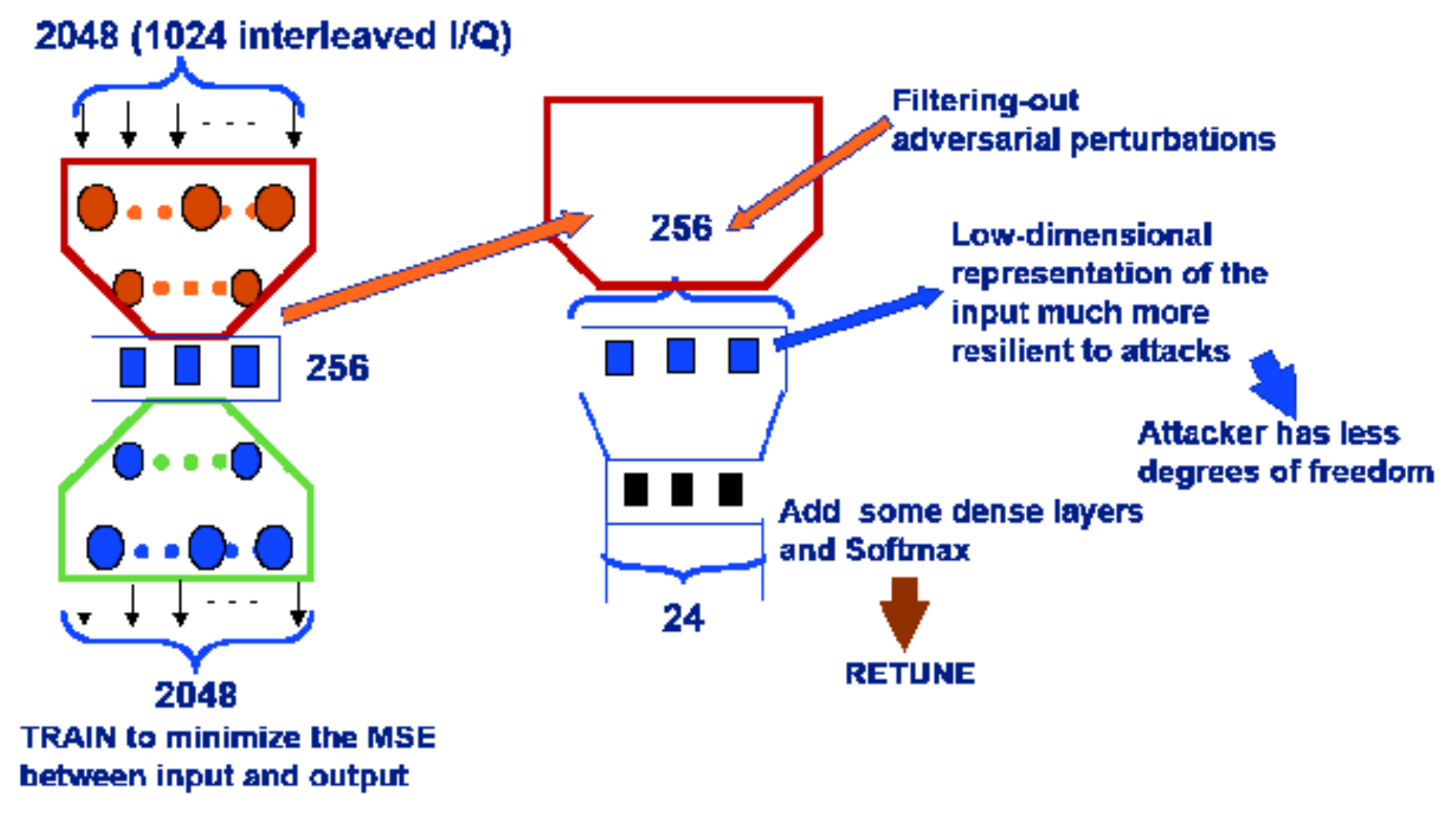}
\caption{{AE-based training}}\vspace{-4mm}   \label{fig:f3}
\end{center}
\end{figure}
%%%%%%%%%%%%%%%%%%%%%%%%%%%%%%%%%%%%%%%%%%%%%%%%%%%%%%%%%%%%%%%
We conducted the AE based training by training the AE$'$s encoder of the same architecture as the CNN presented in Figure~\ref{fig:f2a}, and then transferring the weights to the CNN classifier. The architecture of the AE consisted of such an encoder (red in Figure~\ref{fig:f3}), and the decoder (green in Figure~\ref{fig:f3} ), which is the encoder’s mirror image. Notice from Figure~\ref{fig:f2a} that the encoder consisted of several blocks of 1-D convolutional and max-pooling layers, with layers’ widths narrowing down from 2048 to 256 neurons. Mirroring replaces 1-D convolutions with deconvolutions, and max-pooling with the matching upsampling. The process of training the classifier based on the AE training is shown in Figure~\ref{fig:f3}, where the AE is trained to minimize the mean-square error (MSE) distance between the input and output. 

\begin{figure}[t] %FIGURE 2b
\vspace{-1mm}
\begin{center}
%\hspace{-5mm} \includegraphics [width=2.2in]{unatacked0p1.eps} %sampling.eps 
\hspace{-5mm} \includegraphics [width=2.2in]{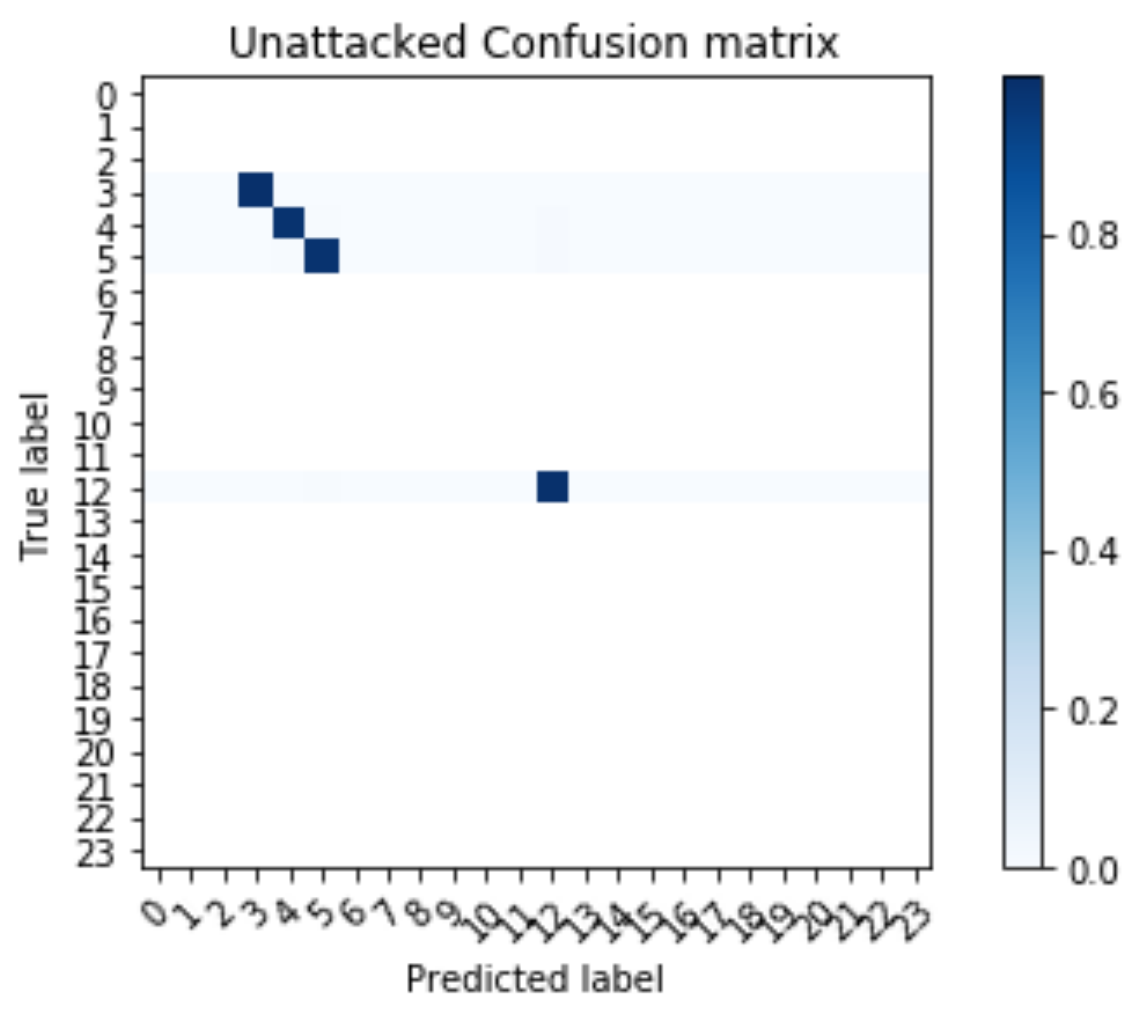}
\caption{{Confusion matrix of unattacked CNN trained on classes 3,4,5 and 12 (BPSK, QPSK, 8-PSK,16QAM)}}\vspace{-4mm}   \label{fig:f2b}
\end{center}
\end{figure}

\begin{figure}[t] %FIGURE 2b
\vspace{-1mm}
\begin{center}
%\hspace{-5mm} \includegraphics [width=2.4in]{NormattConfBud.eps} %sampling.eps 
\hspace{-5mm} \includegraphics [width=2.4in]{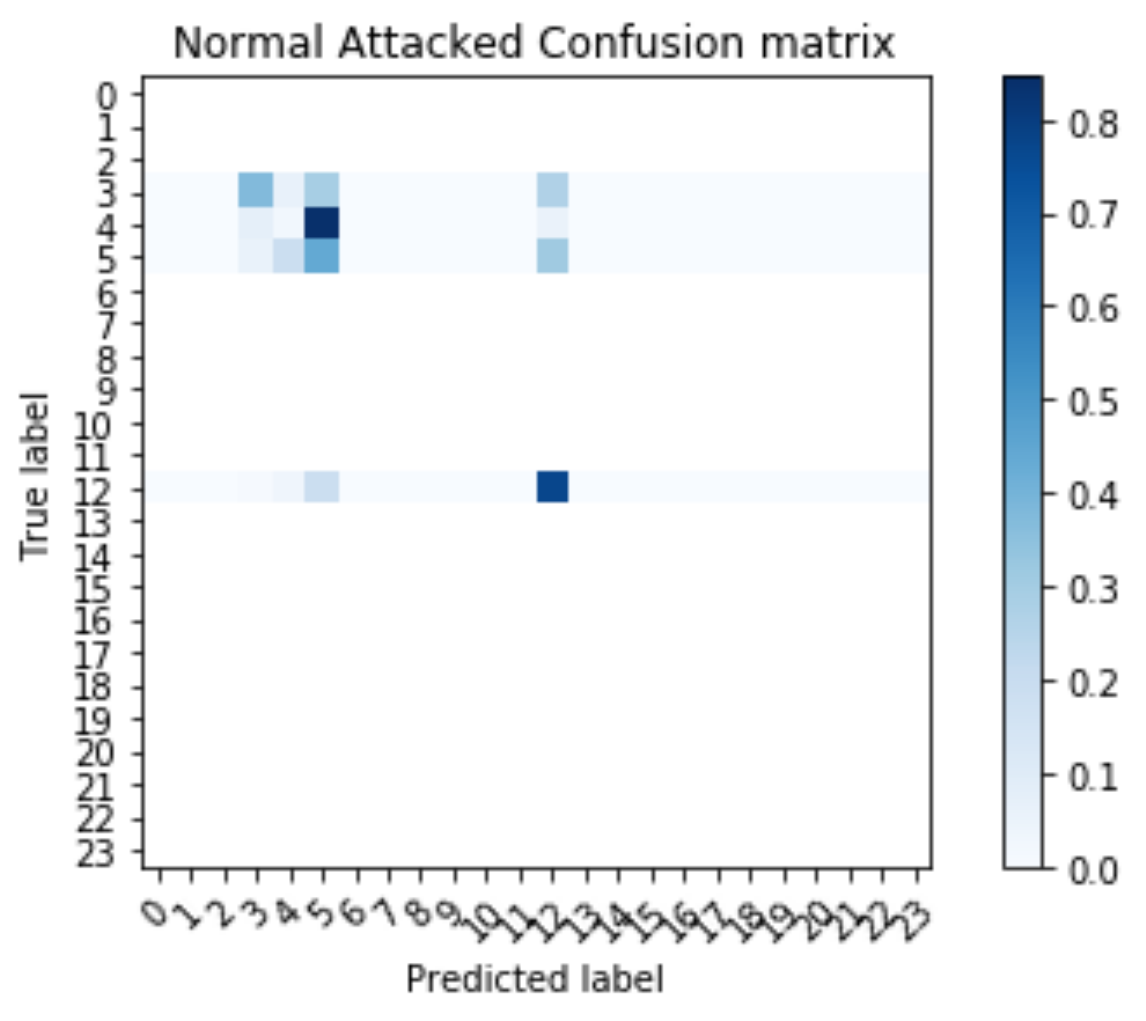}
\caption{{Confusion matrix of the FGSM-attacked classically trained CNN for classes 3,4,5 and 12 (BPSK, QPSK, 8-PSK,16QAM)}}\vspace{-4mm}   \label{fig:f2c}
\end{center}
\end{figure}

\begin{figure}[t] %FIGURE 2b
\vspace{-1mm}
\begin{center}
%\hspace{-5mm} \includegraphics [width=2.4in]{AEattConfBud.eps} %sampling.eps 
\hspace{-5mm} \includegraphics [width=2.4in]{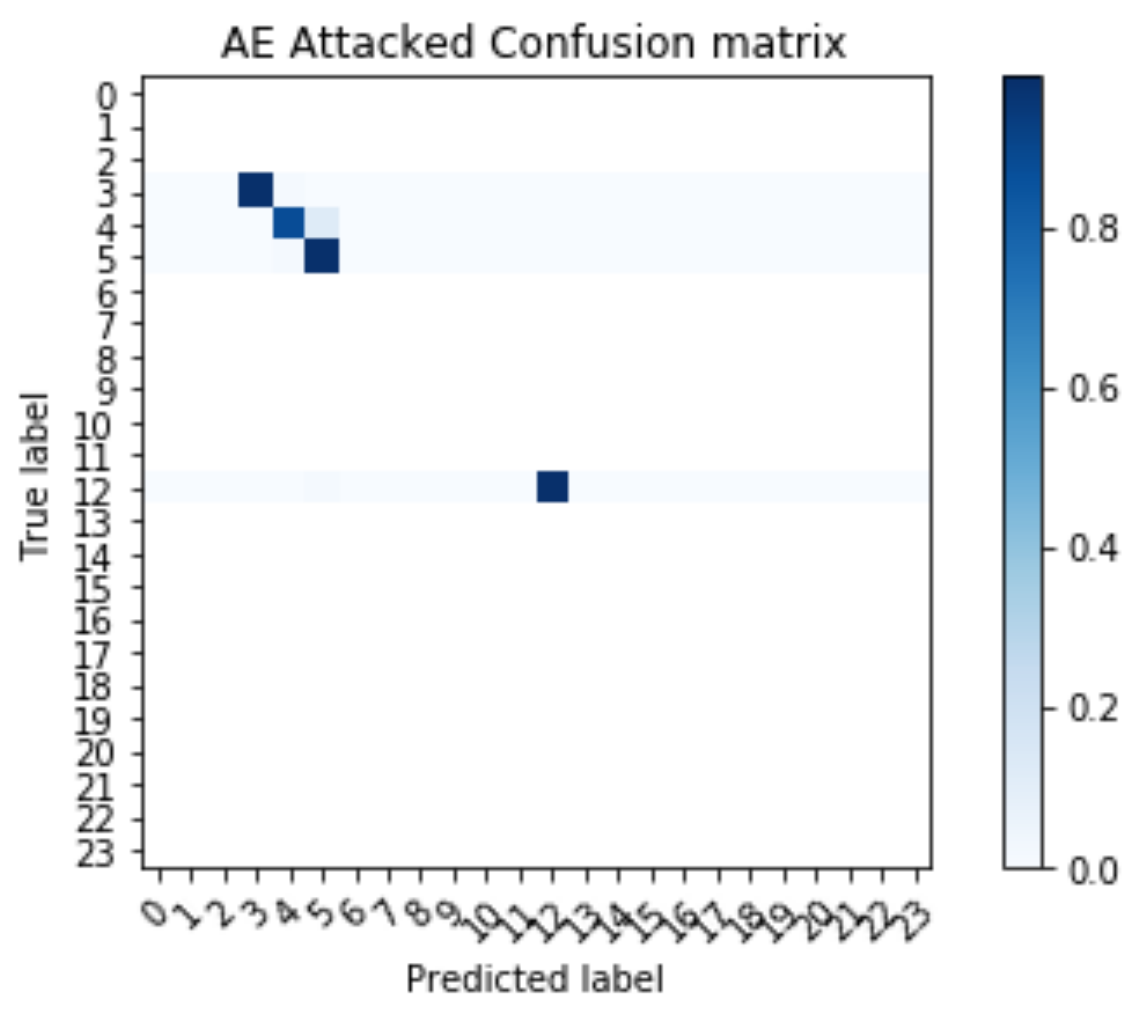}
\caption{{Confusion matrix of FGSM-attacked AE-pretrained CNN trained on classes 3,4,5 and 12 (BPSK, QPSK, 8-PSK,16QAM)}}\vspace{-4mm}   \label{fig:f2d}
\end{center}
\end{figure}

Figure~\ref{fig:f2} plots the classification accuracy of the 4 modulations with and without an FGSM attack ($\epsilon=0.1$), utilizing dashed lines for the AE-trained classifier (red for legitimate data, blue for adversarial). Note that the legitimate accuracy of the AE-trained classifier is fixed, as Figure~\ref{fig:f2} (and Figure~\ref{fig:f2} too) plots the accuracy over the training epochs of the classically trained classifier (i.e., once the AE-based classifier is already trained). Although this is hard to see for the AE-based network, this kind of plotting makes the adversarial accuracy for both DL networks non-constant since adversarial examples are created at each training epoch based on the current loss function. Nevertheless, Figure~\ref{fig:f2} shows how the AE training makes the network more resilient. This is for the FGSM ($\epsilon=0.1$), but similar results are observed for other attack methods. In addition, in Figures~\ref{fig:f2b}~to~\ref{fig:f2d} we present how the accuracy of each of the classified modulations is affected. Figure~\ref{fig:f2b} shows the confusion matrix for the unattacked CNN network, and Figures~\ref{fig:f2c}~and~\ref{fig:f2d} present confusion matrices for the FGSM attacked CNNs, trained conventionally and by AE pretraining, respectively.

\begin{figure}[t] %FIGURE 2
\vspace{-1mm}
\begin{center}
%\hspace{-5mm} \includegraphics [width=3.1in]{acc11NI.eps} %sampling.eps 
\hspace{-5mm} \includegraphics [width=3.1in]{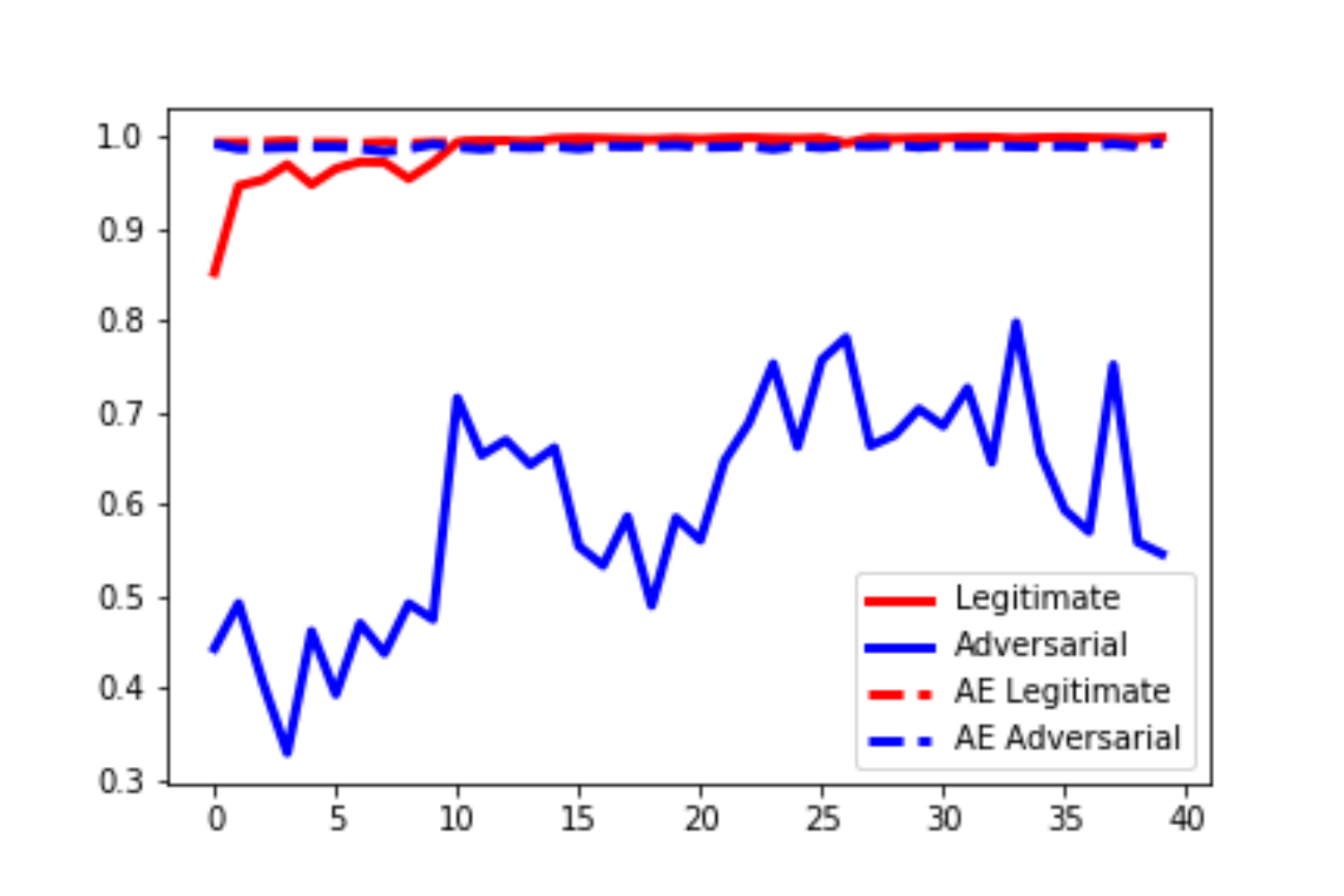}
\caption{{Accuracy (over training epochs) of a 1D-CNN classifier when classically trained, vs pre-trained by autoencoder, with a grey-box FGSM attack}}\vspace{-4mm}   \label{fig:f2g}
\end{center}
\end{figure}

The adversary can only assess the deployed network. He would not know that the encoder layers have not been trained on this network, and it approached the AdEx design in a classical way by utilizing the loss function dependent on the trained weights, which are know to him. This makes his attack a white-box attack by definition, as the adversary knows the network weights, the number and type of layers, the number of convolutional channels and size of convolutional kernels.  An interesting effect is presented in Figure~\ref{fig:f2g} when a grey-box attack is performed, which is the weaker attack when everything else but the actual weights are known. In simpler terms, the adversarial examples are created on an independently classically trained network, but applied to the AE trained network. 
The AE-trained network still maintains resilience against AdExs but the original network performs slightly better for legitimate examples. 
 This promising AE-based defense can be further improved by drawing on our research on channel-robust Stacked Denoising Autoencoder (SDAE).% (skfaddref).
%%%%%%%%%%%%%%%%%%%%%%%%%%%%%%%%%%%%%%%%%%%%%%%%%%%%%%%%%%%%%%
%%%%%%%%%%%%%%%%%%%%%%%%%%%%%%%%%%%%%%%%%%%%%%%%%%%%%%%%%%%%%%%

%%%%%%%%%%%%%%%%%%%%%%%%%%%%%%%%%%%%%%%%%%%%%%%%%%%%%%%%%%%%%%
 %%%%%%%%%%%%%%%%%%%%%%%%%%%%%%%%%%%%%%%%%%%%%%%%%%%%%%%%%%%%%
 
%%%%%%%%%%%%%%%%%%%%%%%%%%%%%%%%%%%%%%%%%%%%%%%%%%%%%%%%%%%%%

%%%%%%%%%%%%%%%%%%%%%%%%%%%%%%%%%%%%%%%%%%%%%%%%%%%%%%%%%%%%%%

%%%%%%%%%%%%%%%%%%%%%%%%%%%%%%%%%%%%%%%%%%%%%%%%%%%%%%%%%%%%%%
\begin{figure}[t] %FIGURE 4
\vspace{-5mm}
\begin{center}
%\hspace{-5mm} \includegraphics [width=3.5in]{probclusters20CNN.eps}\vspace{-4mm}\\ 
\hspace{-5mm} \includegraphics [width=3.5in]{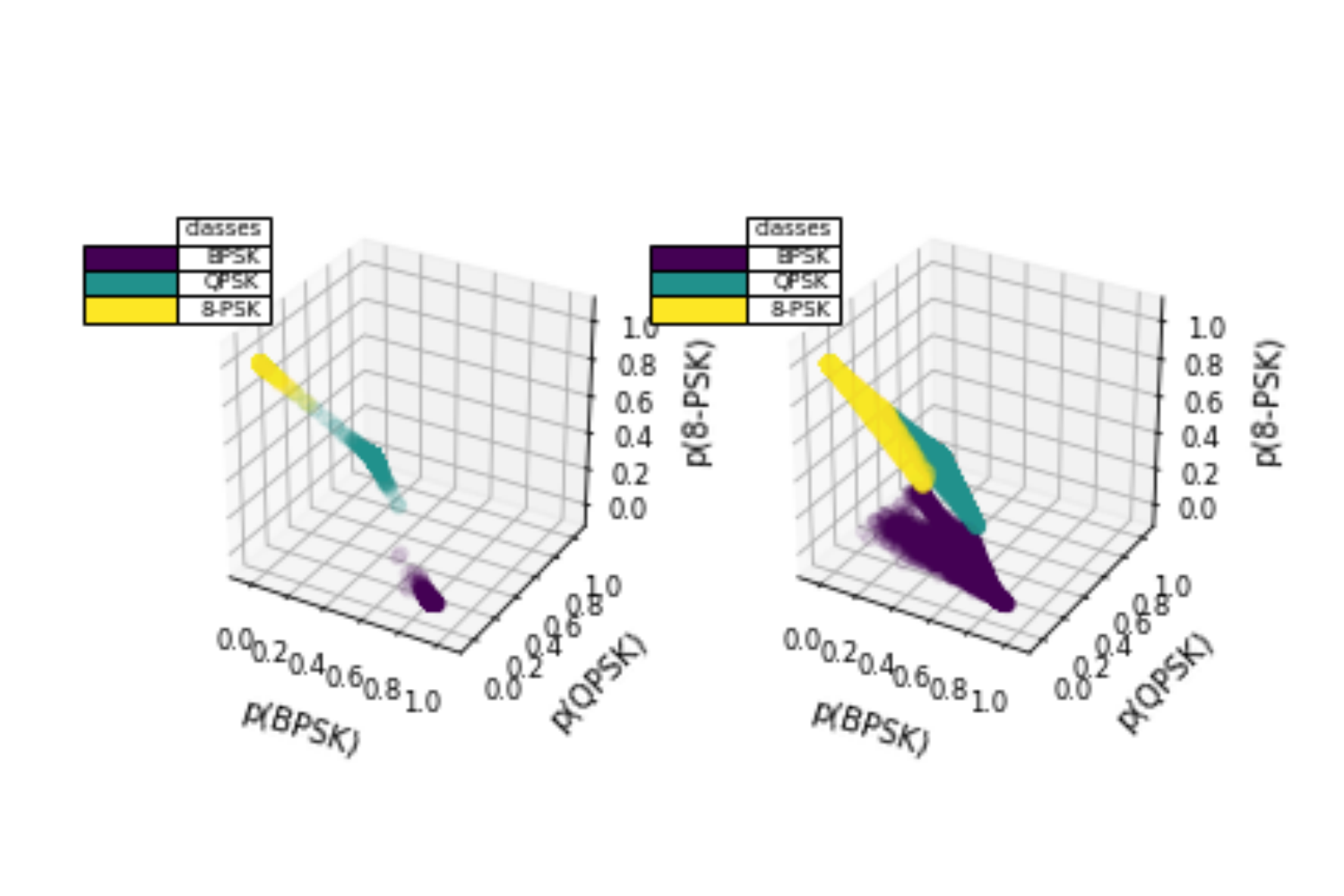}\vspace{-4mm}\\ 
\begin{tabular}{c c}
{\small \textbf{legitimate} \qquad \qquad \qquad}   & {\small \textbf{adversarial}} \vspace{-2mm}
\end{tabular}
\caption{{Classically trained classification of 3 modulations (BPSK, QPSK,8-PSK) shows very different distribution of output probabilities between legitimate and adversarial examples after 40 training epochs}}\vspace{-4mm}   \label{fig:f4}
\end{center}
\end{figure}
%%%%%%%%%%%%%%%%%%%%%%%%%%%%%%%%%%%%%%%%%%%%%%%%%%%%%%%%%%%%%%
%%%%%%%%%%%%%%%%%%%%%%%%%%%%%%%%%%%%%%%%%%%%%%%%%%%%%%%%%%%%%%
\begin{figure}[t] %FIGURE 4
\vspace{-3mm}
\begin{center}
%\hspace{-5mm} \includegraphics [width=3.5in]{probclusters20AE1.eps}\vspace{-4mm}\\ 
\hspace{-5mm} \includegraphics [width=3.5in]{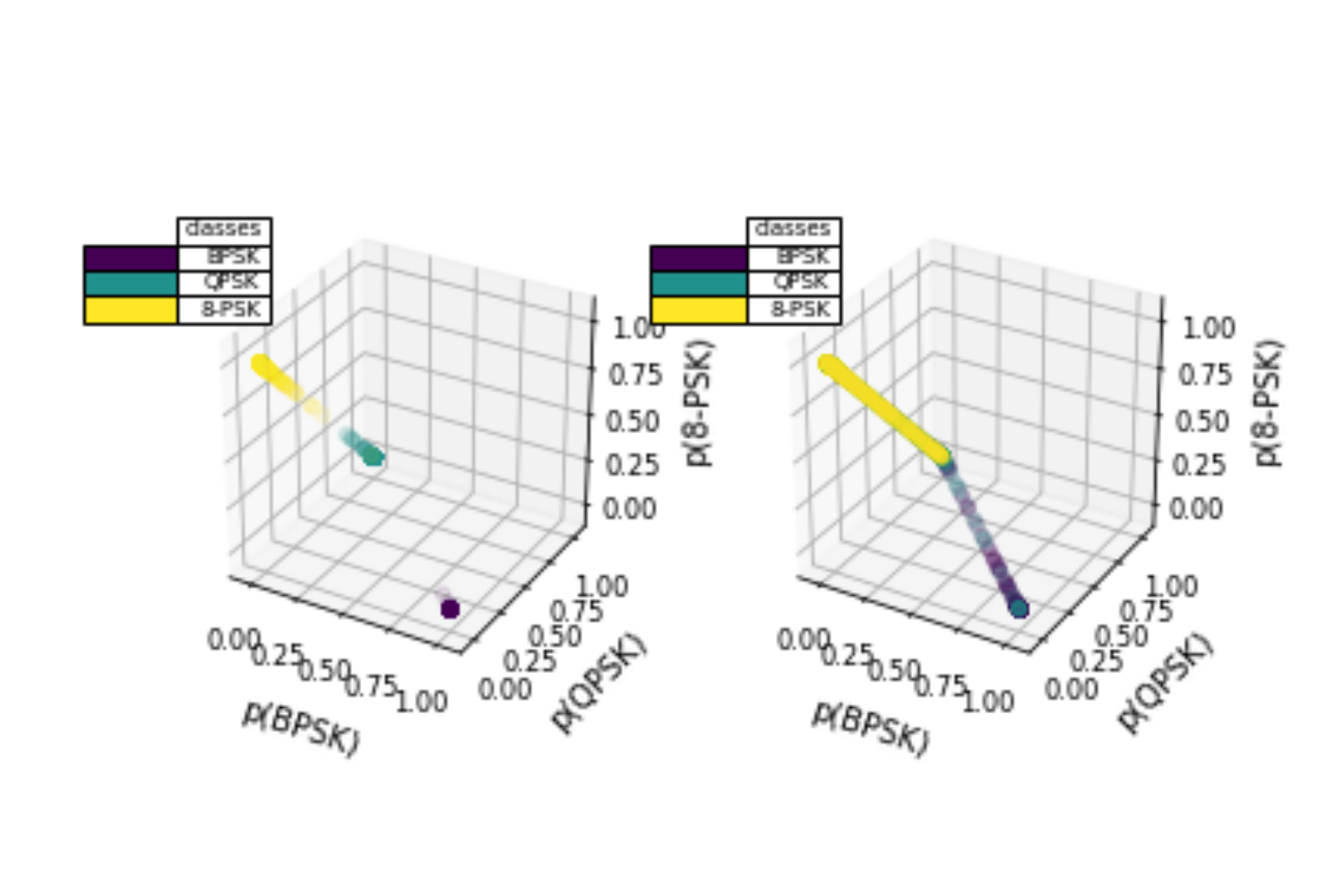}\vspace{-4mm}\\ 
\begin{tabular}{c c}
{\small \textbf{legitimate}\qquad \qquad \qquad}   & {\small \textbf{adversarial}} \vspace{-2mm}
\end{tabular}
\caption{{AE-pretrained classification of 3 modulations (BPSK, QPSK,8-PSK) shows very different distribution of output probabilities between legitimate and adversarial examples after 40 training epochs}}\vspace{-4mm}   \label{fig:f5}
\end{center}
\end{figure}
%%%%%%%%%%%%%%%%%%%%%%%%%%%%%%%%%%%%%%%%%%%%%%%%%%%%%%%%%%%%%%
\vspace{-2mm}
\subsection{How the AE changes the separating hyper-planes}
We have seen from Figures~\ref{fig:f2}, \ref{fig:f2b}, \ref{fig:f2c}, \ref{fig:f2d} that the AE pretraining significantly reduces the effect of FGSM adversarial examples. The question that arises is whether we can combine this kind of mitigation of the attack with the defense methods based on detecting and discarding adversarial examples. To address this we perform a statistical test that compares adversarial and legitimate examples in two experiments: 1)~when the CNN network is not defended, and 2)~when it is defended by the AE-based pretraining. The test that utilizes the output of the Softmax layer is motivated by Figures~\ref{fig:f4}~and~\ref{fig:f5}. Note that we refer to the class probabilities computed by the Softmax layer as the outputs of the classifier:
\begin{align}\eqnlabel{softmax}
P (y = c|x) =\frac{\exp(w_c^T f(x)+b_c)}{\sum_{i\in C}{\exp(w_i f(x)+b_i )}},
\end{align}
where $c \in C, $ $C$ is the set of classes that we perform the inference on, and $w_c$ and $b_c$ are the weight and the bias of the Softmax layer for class $c$. $f(x)$ is the input to that layer for each data point $x$. For the sake of visualization, both figures are based on the 3-class classifier trained on BPSK, QPSK and 8-PSK modulation data points. Hence, the outputs of the Softmax layer are 3-dimensional vectors that are plotted in the figures for all data points utilized for training (close to 20,000, shown on the left), and for their adversarial examples, shown in the plot on the right. The elements of the vectors are values between 0 and 1, representing the probabilities of the classes \eqnref{softmax}. Figure~\ref{fig:f4} shows these vectors after 40 epochs of training the CNN network conventionally, which is upon the convergence of the loss function and after the achieved accuracy exceeded 99\%. Figure~\ref{fig:f5} shows the same after the AE-trained network has converged to its optimal performance.

It is easy to see that the AE training changes the distribution of legitimate, and especially adversarial outputs.
Let us observe first that in Figure ~\ref{fig:f5}-left, representing legitimate examples, the BPSK-classified data points (purple) cluster in the area close to the vertex (1,0,0), denoting the probability of 1 for BPSK and 0 for QPSK and 8-PSK. Similarly, the points classified as QPSK and 8-PSK cluster around the vertices corresponding to probability of 1 for QPSK and 8-PSK, respectively. The left-hand side of Figure~\ref{fig:f4} shows the similar effects for conventionally trained network, although the outputs for each classification are more smeared, which matches the accuracy plot in Figure~\ref{fig:f2}. The right-hand side of Figure~\ref{fig:f4} shows that the output vectors for the adversarial examples of a conventionally trained CNN are distributed across a wide range of values, and the clusterization effect is lost. The same plot on the right-hand side of Figure~\ref{fig:f5} shows less variance in the adversarial outputs, i.e., they are projected along a couple of lines.
\vspace{-2mm}
\subsection{Kolmogorov-Smirnov Test for Output Layer Probabilities}
\begin{figure}[t] %FIGURE 6
\vspace{-1mm}
\begin{center}
%\hspace{+2mm} \includegraphics [width=3.3in]{KStablesAE.eps}\\ 
\hspace{+2mm} \includegraphics [width=3.3in]{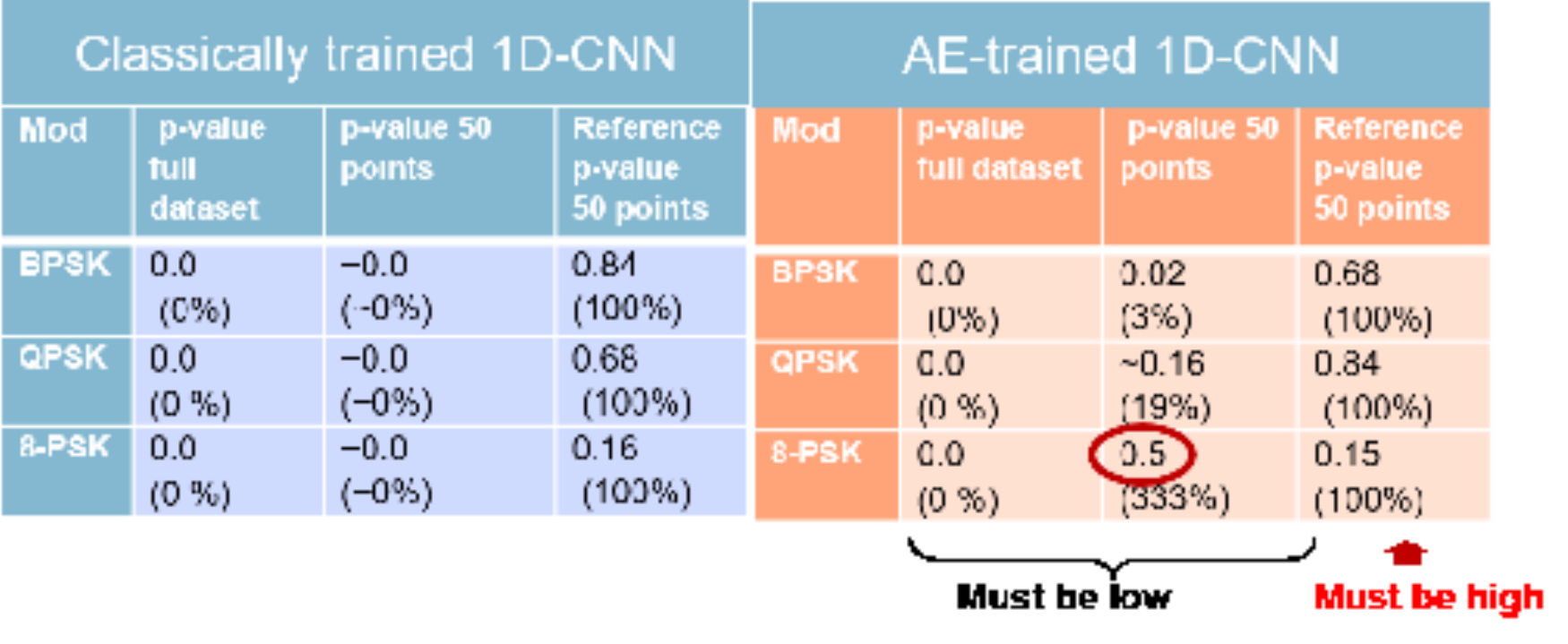}\\
\vspace{-2mm}\begin{tabular}{c c}
\hspace{+2mm} Table 1 & \qquad \qquad \qquad \qquad Table 2
\end{tabular}
\vspace{-2mm}
\caption{{KS-test results, based on output probabilities \eqnref{softmax},  for experiment with classically trained CNN (left), and another with AE-trained CNN (right)}}\vspace{-8mm}   \label{fig:f6}
\end{center}
\end{figure}

This Kolmogorov-Smirnov (KS) two-sample test (see \cite{KSref} and references therein) is performed on the two sets of vector outputs of the classifier. 
We performed each of the tests for the two experiments described above - one with outputs from conventionally trained CNN network, and another with the AE-pretrained outputs. Columns of the tables in Figure~\ref{fig:f6} show for each experiment  the 3 instances of the 2-sample KS test between the outputs: 1) Entire legitimate output dataset per class  vs. entire adversarial output dataset per predicted class; 2) A random set of 50 legitimate output vectors of the same class vs a random set of 50 same-class adversarial output values 3)  Control instance (legitimate to legitimate outputs, per class), with a random set of 50 outputs each. 

The KS test declares the confidence (p-value) that the two sets of statistics are from the same distribution. Tables in Figure~\ref{fig:f6} display those confidence values for the 3 instances described above. Small confidence in the 1st and 2nd column show that AdExs {\em are not drawn from the same distribution as the original data}, and can thus be detected using this test. The 3rd column quantifies the confidence in such a claim, with one being the highest confidence. We see that Table~2 (AE-trained classifier) has much lower confidence in the outcomes of statistical tests, which is expected. This is especially true for small size statistics (50 samples), i.e., if we want to conduct the test in real time, and for a higher order modulation. Obviously, the test design should be more sensitive when used along with mitigation strategies, and/or a belief  network must be used to adapt the decision based on other outcomes. For additional information regarding the KS-based tests aimed to  detect and discard  RF adversarial examples please see out prior work in \cite{PAPRskfrm}. The KS test based on the output probabilities will likely show different results when evaluated at a classifier that is trained on clean RF samples but receives the data points OTA, which causes a different distribution shift due to the channel effects.  We plan to evaluate these effects in future work.
%%%%%%%%%%%%%%%%%%%%%%%%%%%%%%%%%%%%%%%%%%%%%%%%%%%%%%%%%%%%%%%
%%%%%%%%%%%%%%%%%%%%%%%%%%%%%%%%%%%%%%%%%%%%%%%%%%%%%%%%%%%%%%

%\section
%%%%%%%%%%%%%%%%%%%%%%%%%%%%%%%%%%%%%%%%%%%%%%%%%%%%%%%%%%%%%%%
%%%%%%%%%%%%%%%%%%%%%%%%%%%%%%%%%%%%%%%%%%%%%%%%%%%%%%%%%%%%%%
\vspace{-2mm}
\section{Conclusion}%\vspace{-1mm}
We showed that pre-training deep learning classifiers in the RF domain by an autoencoder (AE) mitigates the deceiving effect of adversarial examples (AdExs). The classifier that we designed for evaluation of this defense method is based on several 1-dimensional convolutional and max-pooling layers, and two regularized dense layers at the bottom of the network. The classification accuracy of the trained network was satisfactory on the legitimate dataset, which consists of four differently modulated RF signals. Despite the improvements due to AE-based pretraining, there is some residual decrease in the accuracy of the attacked classifier that should be addressed by different methods. We intend to address this in our future work by expanding the AE-based defense to a denoising AE, which is also likely to increase its robustness against the receiver noise, i.e., unintentional input corruption. We also explored if we can combine this kind of mitigation of the attack with the defense methods based on detecting and discarding adversarial examples. We show that detection methods based on the statistical tests to detect a distribution shift of the values at the output of the DL classifier are not as effective as when applied to an undefended classifier. This should be considered when the detection and mitigation by pretraining are combined to strengthen the classifier$'$ robustness to adversarial attacks.  The validity of the proposed defense should be verified in terms of robustness to corruption incurred at the receiver due to over-the-air delivery of RF data points, which we plan to evaluate in future work.%The robustness of the attacks to deep learning via adversarial examples, and of the defense mechanisms against them is an open problem, and this paper is an attempt to address its manifestations and specificities in the RF domain.  
\vspace{-2mm} 
\bibliographystyle{IEEEtran}%
\bibliography{adversarial}
\end{document}